\documentclass[final,5p,times,twocolumn]{elsarticle}

\usepackage{graphicx}

\usepackage{amssymb}
\usepackage{amsthm}

\usepackage{bm}
\usepackage{here}
\usepackage{color}

\definecolor{mygreen}{rgb}{0.2,0.8,0.2}

\begin{document}

\begin{frontmatter}


\title{Wavepacket Dynamics in One-Dimensional System with 
Long-Range Correlated Disorder
}




 \author{Hiroaki S. Yamada}
\address{Yamada Physics Research Laboratory, Aoyama 5-7-14-205, Niigata 950-2002, Japan}

\begin{abstract}
We numerically investigate dynamical   property in the one-dimensional 
tight-binding model with long-range correlated disorder having power spectrum
 $1/f^\alpha$ ($\alpha:$spectrum exponent) generated by Fourier filtering method.
For relatively small $\alpha<\alpha_c(=2)$ time-dependence of mean square 
displacement (MSD) of the initially localized wavepacket shows ballistic spread
 and localizes as time elapses. It is shown that  $\alpha-$dependence of  
the dynamical localization length (DLL) determined by the MSD exhibits a simple 
scaling law in the localization regime  for the relatively weak disorder strength $W$.
Furthermore, scaled MSD by the DLL almost obeys an universal function 
from the ballistic to the localization regime in the various combinations of 
the parameters $\alpha$ and  $W$.
\end{abstract}


\end{frontmatter}


\def\ni{\noindent}
\def\nn{\nonumber}
\def\bH{\begin{Huge}}
\def\eH{\end{Huge}}
\def\bL{\begin{Large}}
\def\eL{\end{Large}}
\def\bl{\begin{large}}
\def\el{\end{large}}
\def\beq{\begin{eqnarray}}
\def\eeq{\end{eqnarray}}

\def\eps{\epsilon}
\def\th{\theta}
\def\del{\delta}
\def\omg{\omega}

\def\e{{\rm e}}
\def\exp{{\rm exp}}
\def\arg{{\rm arg}}
\def\Im{{\rm Im}}
\def\Re{{\rm Re}}

\def\sup{\supset}
\def\sub{\subset}
\def\a{\cap}
\def\u{\cup}
\def\bks{\backslash}

\def\ovl{\overline}
\def\unl{\underline}

\def\rar{\rightarrow}
\def\Rar{\Rightarrow}
\def\lar{\leftarrow}
\def\Lar{\Leftarrow}
\def\bar{\leftrightarrow}
\def\Bar{\Leftrightarrow}

\def\pr{\partial}

\def\Bstar{\bL $\star$ \eL}
\def\etath{\eta_{th}}
\def\irrev{{\mathcal R}}
\def\e{{\rm e}}
\def\noise{n}
\def\hatp{\hat{p}}
\def\hatq{\hat{q}}
\def\hatU{\hat{U}}

\def\iset{\mathcal{I}}
\def\fset{\mathcal{F}}
\def\pr{\partial}
\def\traj{\ell}
\def\eps{\epsilon}




\section{Introduction}
In one-dimensional 
disordered systems (1DDS) with uncorrelated disorder 
all eigenstates are exponentially localized and spectrum is pure point
\cite{ishii73,abrahams79}.
In one-dimensional tight-binding model (1DTBM) 
it has been numerically known that  correlations with power spectrum
$S(f) \sim 1/f^\alpha$($\alpha \geq 2$)
 generated  by Fourier filtering method (FFM)
 in the on-site potential  
 delocalize the eigenstates and induce localization-delocalization transition (LDT), 
where $f$ denotes frequency and $\alpha$ is spectrum exponent controlling 
the spatial correlation
\cite{moura98,zhang02,shima04,kaya07,gong10,garcia09,iomin09,izrailev12,croy11,deng12,albrecht12,petersen13}.
The LDT also depends on the potential strength $W$ controlling the distribution
width of the disorder of on-site energy in the 1DTBM.
The critical spectrum exponent separating
 the localized and delocalized states 
$\alpha_c(=2)$ for the relatively small $W$ scaled by transfer energy.
However, the quantum diffusion in the non-stationary region, $1<\alpha<2$, 
hardly been investigated so far 
\cite{yamada92,oliveira01,sales12,cheraghchi05,lazoa10,yamada15,pinto05,santos06,wells08,diaz09,assuncao11}
Note that the localization of wavepacket dynamics
 (dynamical localization) is sufficient condition for the pure point spectrum.

In this paper, we focus only on localization phenomena and 
the characteristics of the localization length in the wavepacket dynamics 
for $\alpha<\alpha_c$ and relatively large $W(\geq 0.5)$.


We report the characteristic $\alpha/W-$dependences of 
the dynamical localization length (DLL) of 
 the quantum diffusion of the initially localized wavepacket 
in 1DTBM with long-range correlated disorder
generated by FFM.
It should be noted that we deal with parameter regime that 
the wavepaclet perfectly localized within the accessible numerical calculation.
It seems that the time dependence of MSD of the wavepacket 
changes from ballistic ($ \sim t^2$) to localized ($ \sim t^0$) 
as time $t$ elapses when the potential strength $W$ is relatively large
and at least for $\alpha <\alpha_c$.
It is suggested that the DLL is given as a simple
function of the spectrum index $\alpha$ and the potential strength $W$.
Furthermore, the scaled MSD almost obeys an universal function 
from the ballistic to the localization regime in the various combinations of 
the parameters $(W, \alpha)$.

This paper is organized as follows.
In the next section, we give the 1DDS with the FFM potential.
Sect.\ref{sec:quantum-diffusion} reports about the 
localization behaviour of the initially localized wavepacket.
Summary and discussion  are presented in the last section.

\section{Model}
\label{sec:model}
We examine the dynamical property  
of the initially localized wave packet  $\phi(n,t=0)=\delta_{n,n_0}$ by 
the quantum time-evolution; 
\beq
i\hbar \frac{\pr \phi(n,t)}{\pr t}= \phi(n+1,t)+\phi(n-1,t) + Wv(n)\phi(n,t),
\eeq
where $ n=1,2,...,N$,  and $\hbar=1$ through this calculation.
We characterize the spread of the wavepacket by 
the mean square displacement (MSD),
\begin{eqnarray}
m_2(t) = \sum_{n}(n-n_0)^2 \left< |\phi(n,t)|^2 \right>,
\end{eqnarray}
where $\langle...\rangle$ indicates the ensemble average over different disorder.
In general, in the long-time limit the time-dependence of MSD can be characterized 
by  diffusion exponent $\sigma$ as follows;
\beq
   m_2 \sim t^\sigma.
\eeq
$\sigma=0$ corresponds to the localization, 
$0< \sigma <1$ to subdiffusion,
$\sigma =1$ to normal diffusion,
$1< \sigma <2$ to superdiffusion,
and $\sigma =2$ to ballistic motion.
Also,  $\sigma=2/3$ should correspond to the metal-insulator transition in 3DDS,
whereas $\sigma =1$ and $\sigma =2$ might also be found in 1DDS 
with the stochastically fluctuation and in the periodic systems, respectively.

Disorder of the potential with the long-range correlation is generated by 
the FFM in the following form \cite{moura98,santos06,albrecht12}:
\beq
v(n) = C(\alpha,L) \sum_{k=1}^{L/2} 
\left(\frac{2\pi k}{L}\right)^{-\alpha/2}
\cos\left(\frac{2\pi k n}{L} +\varphi_k \right),
\label{eq:wei-pot}
\eeq
where $\alpha$ is a spectrum index  ($\alpha>0$) related 
the power spectrum
$S(f) \sim 1/f^\alpha$($\alpha \geq 0$) of the sequence $\{v(n) \}$.
 $\{ \varphi_k \}_{k=0}^{L/2}$ are random independent variables chosen in the interval $[0,2\pi]$.
$C(\alpha,L)$ is the normalization constant which is determined 
by a condition $\sqrt{<v(n)^2>-<v(n)>^2}=1$.
We mainly used $L=10^5$, and systemsize $N=2^{12} \sim 2^{14}$ 
under a condition $L>N$ that does not alter qalitative numerical result.
The typical ensemble size used 
here is $100$.
The robustness of the numerical 
calculations has been confirmed in each case.



With this potential sequence, the analysis of the Lyapunov exponent 
have suggested that the LDT (Anderson transition) 
 for the relatively small $W$  occurs in $\alpha_c=2$ \cite{shima04,kaya07}.
In the following we will use $\alpha_c=2$ as a guideline, 
and deal with a regime $\alpha<\alpha_c$ only 
to obtain the numerically accurate DLL.


\section{Wavepacket dynamics and Localization}
\label{sec:quantum-diffusion}
\subsection{spread of wavepacket}
We numerically solved equation (1) to investigate 
$\alpha/W-$dependence of the DLL.

\begin{figure}[htbp]
\begin{center}
\includegraphics[width=6.0cm]{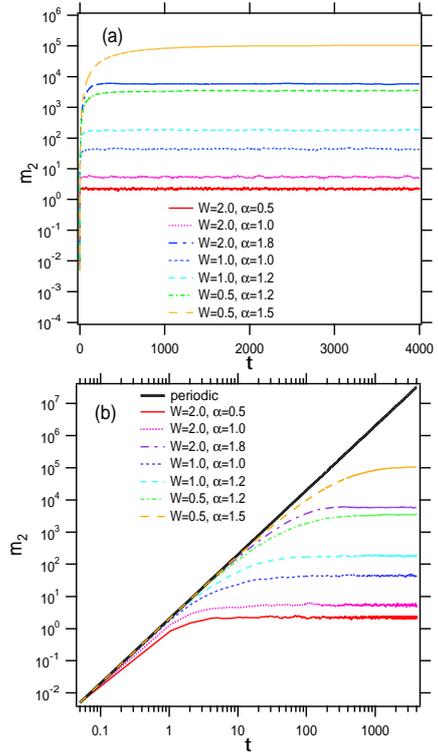}
\caption{
(Color online)
(a)The second moment $m_2$ as a function of time for some values of 
$\alpha$ and $W$.
Panel (b) is the double-logarithmic plot.
The ballistic increase ($m_2 \sim t^2$) of a periodic system
($v_n=1, n=1,2,...,N$) is also plotted by 
black bold line as a reference.
$N=2^{13}$
}
\label{fig:msd-fig1}
\end{center}
\end{figure}

Figure \ref{fig:msd-fig1}(a) shows the time-dependence of MSD 
 for various parameter sets $(W, \alpha)$.
The results in the case of a periodic sequence ($v_n=1, n=1,2,...,N$) 
showing complete ballistic motion ( $m_2 \propto t^2$) are also shown 
as a reference in Fig.\ref{fig:msd-fig1}(b).
It seems that  
the wavepacket dynamically localizes at $t \to \infty$.
We may note that the MSD  evolves 
from the ballistic-like rise ($m_2 \sim t^2$) 
to the complete localization  ($m_2 \sim t^0$),  
after passing through the intermediate spreading regime, 
as seen in the double-logarithmic plot in Fig.\ref{fig:msd-fig1}(b).
The spread is larger along with the rise in $\alpha$ 
if the potential strength $W$ is constant.
Accordingly, the DLL becomes larger values 
as the spectrum exponent $\alpha$ becomes larger, and 
potential strength $W$ becomes smaller.





\subsection{Localization length}
\begin{figure}[htbp]
\begin{center}
\includegraphics[width=8.0cm]{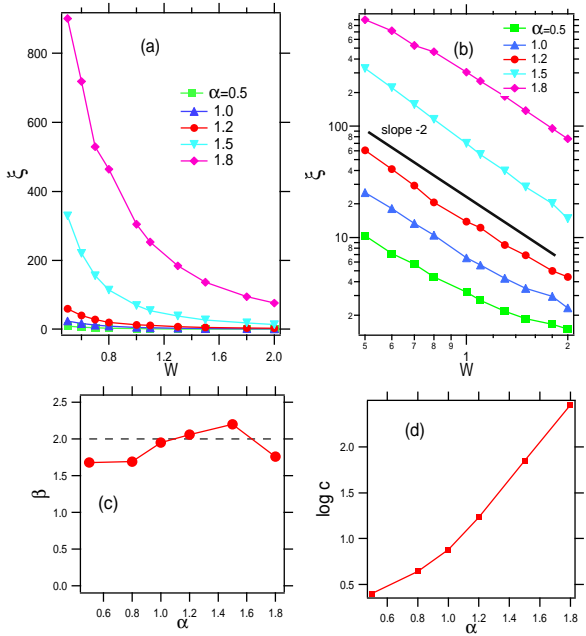}
\caption{
(Color online)
(a)Dynamical localization length $\xi$ as a function of the potential strength $W$
for some values of $\alpha$.
$N=2^{13}$.
(b) The double logarithmic plot.
Panels (c) and  (d) are the slope and segment as a function of $\alpha$
determined by least square fit for the data in (b), respectively.
Note that we can obtain a value $\beta \simeq 2$
by the fit for the smaller $W$ region.
}
\label{fig:LL-alpha-A}
\end{center}
\end{figure}

In general, 
the localization length becomes definite only for the perfect localization ($\sigma=0$),
and infinity for $\sigma>0$,  in a limit $t \to \infty$.
Here, we consider the localization length in the case where 
the localization can be judged clearly from numerical data excluding the size effect, 
as seen in Figs. \ref{fig:msd-fig1}.
We define the {\it dynamical localization length} $\xi$ by MSD as 
$m_2(t \to \infty)=\xi^2$.
Then numerically, we use the time-averaged value in the region where 
the ballistic motion finishes and the value of $m_2$ is stable enough.
Figure \ref{fig:LL-alpha-A} (a) shows the $W-$dependence of 
the $\xi$ numerically.
As seen in Fig \ref{fig:LL-alpha-A} (b), 
if $\alpha$ is the same, there is a tendency to decay 
according to the inverse power-law with respect to $W$.
Therefore, we suppose this functional form is
\beq
  \xi =\frac{c}{W^\beta}.
\eeq
The $\alpha-$dependence of the positive coefficients $c$ and $\beta$ are
 determined numerically by the least squares method.
The result is given in Fig.\ref{fig:LL-alpha-A} (c) and (d), respectively.
We can see that if $ W $ is relatively small, regardless of $\alpha$, 
the power exponent $\beta$ is roughly 2, and the coefficient $c$ 
increases exponentially with respect to $\alpha$.
As a result,  the result suggests the $\alpha/W-$dependence
of the dynamical localization length for $1<\alpha <<\alpha_c$ as follows; 
\beq
  \xi \simeq \frac{\e^{k\alpha}}{W^2}, 
\label{eq:loc-scale}
\eeq
where $k$ is a positive coefficient determined by the slope of (b).
It is consistent with the $W-$dependence,  $\xi \sim W^{-2}$,  
usually seen in 
the 1DDS without correlation (corresponding to $\alpha=0$)
and with the short-range correlation　for the weak disorder limit \cite{sales12}.

\begin{figure}[htbp]
\begin{center}
\includegraphics[width=6.0cm]{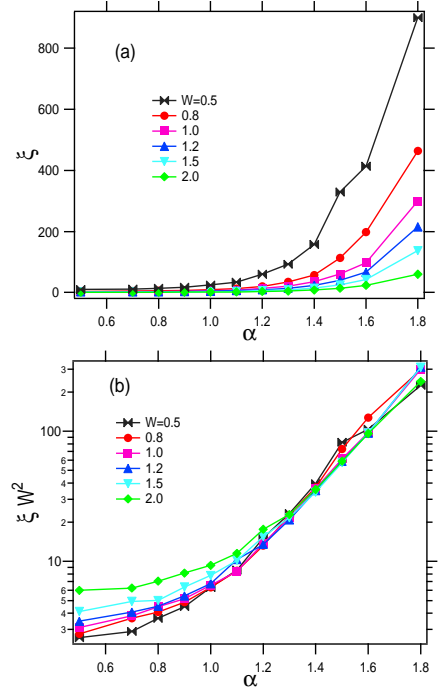}
\caption{
(Color online)
(a)Dynamical localization length $\xi$ as a function of the spectrum index $\alpha$
for some $W$s,  which are determined by ensemble-average and time-average
 for the stable fluctuation regime. $N=2^{13}$.
(b) $\xi W^2$ as a function of the  $\alpha$
for the data in the panel (a).
}
\label{fig:LL-alpha-B}
\end{center}
\end{figure}

Figure \ref{fig:LL-alpha-B} (a) shows the $\alpha-$dependence of 
the DLL $\xi$ numerically obtained 
as a time-averaged value 
from the data  in a long time.
It turns out to be increased sharply towards $\alpha_c=2$ 
as disorder strength $W$ becoms smaller. 
%
Figure \ref{fig:LL-alpha-B} (b) 
shows plots for $\xi W^2$ as a function of $\alpha$ to scale changes due to $W$
 based on the expression (\ref{eq:loc-scale}).
For relatively large $W$, it seems that 
the $\alpha-$dependence of the DLL is roughly scaled 
by taking a single curve regardless of $W$ in this range $1<\alpha<\alpha_c(=2)$.
As we can expect, is has been confirmed that the functional form 
is similar to one given in Fig.\ref{fig:LL-alpha-A}(d).




\subsection{Scaling of Localization Dynamics}
\label{sec:scaling}
\begin{figure}[htbp]
\begin{center}
\includegraphics[width=9.0cm]{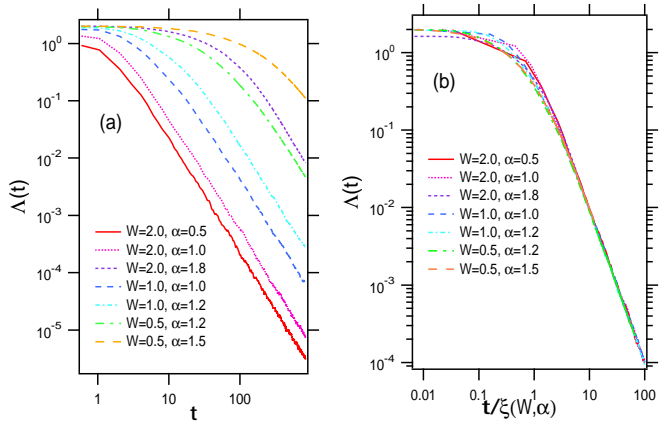}
\caption{(Color online)
(a)Scaled MSD $\Lambda(t)$ as a function of time 
for several sets of the parameters $(W, \alpha)$.
(b) $\Lambda(t)$ as a function of the scaled time
$t/\xi(W,\alpha)$, where $\xi(W,\alpha)$ are determined by MSD.
Note that all axes are in logarithmic scale.
}
\label{fig:lambda-1}
\end{center}
\end{figure}

In this subsection, we investigate the localization characteristics
of the time-dependence of MSD from the ballistic  
to the localization when $W$ is relatively small 
and  $\alpha <\alpha_c$.
We assume that at the critical state the MSD exhibits a ballistic spread
characterized by the
diffusion exponent $\sigma=2$.
In this case, instead of the MSD, we use
the scaled MSD
\begin{eqnarray}
\Lambda(t)\equiv\frac{m_{2}(t)}{t^2}
\label{eq:scale-1}
\end{eqnarray}
with respect to the critical behaviour $t^2$.
This type of the scaling analysis has been performed 
to investigate LDT phenomena at the critical point
for polychromatically perturbed disordered systems \cite{delande07}.
Fig.\ref{fig:lambda-1}(a) shows its temporal evolution of $\Lambda(t)$
at various $W$s and $\alpha$s.
Indeed, the ballistic spread of the wavepacket 
has been observed for $\alpha>\alpha_c$ \cite{santos06}.
Figure \ref{fig:lambda-1}(b) shows $\Lambda(t)$ as a function of 
the scaled time $t/\xi(W,\alpha)$ to characterize the  localization phenomenon.
Here, $W/\alpha-$dependent localization length $\xi (W,\alpha) $ 
is determined by the numerical data of MSD.
It follows that it roughly overlaps from the ballistic to the localized region.
These result suggests that the localization process 
from $m_2 \sim t^2$ to $m_2 \sim t^0$ is scaled by one-parameter 
when the localization occurs.




\section{Summary and discussion}
\label{sect:summary}
In summary, we have numerically studied the nature of  localized property 
of the initially localized wavepacket
 in 1DTBM with long-range correlation generated by FFM.
As a result, we have found that in the localization region 
with small spectrum exponent $\alpha<\alpha_c$ and/or relatively strong
disorder $W>0.5$ the time-dependence of MSD changes 
 from ballistic to localized behaviour as the time elapses.
The dynamical localization length determined by the MSD has been well scaled 
by  Eq.(\ref{eq:loc-scale}) in the parameter regions.
In addition, it has been supported that the time-dependent localization
process can be also scaled by the dynamical localization length.

According to the self-consistent theory of the localization, it is well known that the critical diffusion at the 
localization-delocalization transition point in the $d-$dimensional random systems
 is $m_2 \sim t^{2/d}~(d\geq 3)$ \cite{wolfle10}.
 If we apply this result formally to $d=1$, then the critical diffusion will be 
$m_2\sim t^{2}$  at the transition point 
from the localization to the delocalization in the 1DDS. 
The above numerical results 
support this hypothesis.

\section*{Acknowledgments}
The author would like to thank Professor M. Goda for discussion 
about the correlation-induced delocalization at 
early stage of this study, 
and  Professor E.B. Starikov for proof reading of the manuscript.
The author also would like to acknowledge the hospitality of 
the Physics Division of The Nippon Dental University at Niigata
for  my stay, where part of this work was completed.
The sole author had responsibility for all parts of the manuscript.



\begin{thebibliography}{00}



\bibitem{ishii73} K. Ishii, Prog. Theor. Phys. Suppl. {\bf 53},
77(1973).

\bibitem{abrahams79}
E. Abrahams, P. W. Anderson, D. C. Licciardello, and T. V. Ramakrishnan, 
Rhys. Rev.Lett. {\bf 42}, 673 (1979).



\bibitem{moura98}
F.A.B.F.de Moura, and M.L.Lyra, 
Phys. Rev. Lett. {\bf 81}, 3735(1998).



\bibitem{zhang02}
G.-P. Zhang and S.-J. Xionga, 
Eur. Phys. J. B {\bf 29}, 491-495(2002).

\bibitem{shima04}
H.Shima, T.Nomura and T.Nakayama,
Phys. Rev. B {\bf 70}, 075116(2004).

\bibitem{kaya07}
T. Kaya,
Eur. Phys. J. B {\bf 55}, 49(2007).




\bibitem{gong10}
L.Y. Gong,  P.Q. Tong, and Z.C. Zhou,
Eur. Phys. J. B {\bf 77}, 413-417(2010).

\bibitem{garcia09}
A. M. Garcia-Garcia, and E. Cuevas,
Phys. Rev. B {\bf 79}, 073104 (2009).

\bibitem{iomin09}
A. Iomin,
Phys. Rev. E {\bf 79}, 062102(2009).

\bibitem{izrailev12}
F. M. Izrailev, A. A. Krokhin, and N. M. Makarov, 
Phys. Rep.  {\bf 512}, 125 (2012).



\bibitem{croy11}
A. Croy, P. Cain, and M. Schreiber, 
Eur. Phys. J. B {\bf 82}, 107 (2011).

\bibitem{deng12}
Chao-Sheng Deng, and HuiXu,
Physica E {\bf 44} 1473-1477(2012).



\bibitem{albrecht12}
C. Albrecht and S. Wimberger,
Phys. Rev. B {\bf 85}, 045107 (2012).


\bibitem{petersen13}
G.M. Petersen and N. Sandler,
Phys. Rev. B {\bf 87}, 195443 (2013).
\bibitem{yamada92}
H. Yamada,  M. Goda,  Y.  Aizawa, and  M. Sano,
J. Phys. Soc. Jpn. {\bf 61},  3050-3053 (1992).

\bibitem{oliveira01}
C.R. de Oliveira and G.Q. Pellegrino, 
J. Phys. A {\bf 34}, L239-L243 (2001).  

\bibitem{sales12}
M. O. Sales,  F. A. B. F. de Moura, 
Physica E {\bf 45}, 97-102(2012).

\bibitem{cheraghchi05}
H. Cheraghchi, S. M. Fazeli, and K. Esfarjani, 
Phys. Rev. B {\bf 72}, 174207(2005).






\bibitem{lazoa10}
E. Lazoa, E. Diezb, 
Phys. Lett. A {\bf 374},  3538-3545(2010).
















\bibitem{yamada15}
H.S. Yamada,
Eur. Phys. J. B {\bf 88}, 264 (2015).













\bibitem{pinto05}
R. A. Pinto, M. Rodriguez, J. A. Gonzalez, and E. Medina,
Phys. Rev. A {\bf 341}, 101-106(2005).









\bibitem{santos06}
B.Santos, L.P.Viana,M.L.Lyra, F.A.B.F. de Moura,
Solid State Commum. {\bf 138}, 1-5(2006). 

\bibitem{wells08}
P.R.Wells Jr., J. dAlbuquerque e Castro, and S.L.A. de Queiroz,
Phys. Rev. B {\bf 78}, 035102(2008).  


\bibitem{diaz09}
E. Diaz,  and F. Domínguez-Adame,
Chem. Phys. {\bf 365}, 24–29(2009).

\bibitem{assuncao11}
T.F. Assunçao, M.L. Lyra, F.A.B.F. de Mour, and F. Domínguez-Adame,
Phys. Lett. A {\bf 375}, 1048–1052(2011).












\bibitem{delande07}
R. C. Kuhn, O. Sigwarth, C. Miniatura, D. Delande, and C. A. Muller, 
New J. Phys. {\bf 9}, 161 (2007).

\bibitem{wolfle10}
P. Wölfle and D. Vollhardt, 
Int. J. Mod. Phys B 24, 1526–1554 (2010).

























\end{thebibliography}


\end{document}